\documentstyle[prd,aps]{revtex}

\newcommand{\bea}{\begin{eqnarray}}
\newcommand{\eea}{\end{eqnarray}}

\begin{document}

%%%%%%%%%%%%%%%%%%%%%%%%%%%%%%%%%%%%%%%%%%%%%%%%%%%%%%%%%%%%%%%
\draft
\twocolumn[\hsize\textwidth\columnwidth\hsize\csname
@twocolumnfalse\endcsname

%%%%%%%%%%%%%%%%%%%%%%%%%%%%%%%%%%%%%%%%%%%%%%%%%%%%%%%%%%%%%%%
\title{Roles of a coherent scalar field on the evolution of cosmic structures}
\author{Jai-chan Hwang}
\address{Department of Astronomy and Atmospheric Sciences,
         Kyungpook National University, Taegu, Korea}
\date{\today}
\maketitle

%%%%%%%%%%%%%%%%%%%%%%%%%%%%%%%%%%%%%%%%%%%%%%%%%%%%%%%%%%%%%%%
\begin{abstract}

A coherently oscillating scalar field, an axion as an example, is known to 
behave as a cold dark matter.
The arguments were usually made in the Newtonian context.
Ratra proved the case in relativistic context using the synchronous gauge.
In this paper we present another proof based on a more suitable 
gauge choice, the uniform-curvature gauge, which fits the problem.
By a proper time averaging the perturbed oscillating scalar field
behaves as a cold dark matter on the relevant scales including the
superhorizon scale.

\end{abstract}

\noindent

%%%%%%%%%%%%%%%%%%%%%%%%%%%%%%%%%%%%%%%%%%%%%%%%%%%%%%%%%%%%%%%
\vskip2pc]
%%%%%%%%%%%%%%%%%%%%%%%%%%%%%%%%%%%%%%%%%%%%%%%%%%%%%%%%%%%%%%%
{\it 1. Introduction:}
There exists a large amount of literature concerning the role of a
coherently oscillating scalar particle as a cold dark matter 
\cite{Axion-CDM,Brandenberger,Sasaki,Ratra}. 
Most of the arguments were based on the Newtonian or
qualitative analyses \cite{Axion-CDM}. 
A relativistic treatment can be found in \cite{Brandenberger}, however,
there also exist the mixed statements concerning the role \cite{Sasaki}.
B. Ratra presented a fully relativistic analyses based on 
the synchronous gauge choice \cite{Ratra}.
In the synchronous gauge the analysis becomes unnecessarily complicated 
and one has to be careful of tracing the remaining gauge mode
as the author of \cite{Ratra} did.
In this paper we would like to derive the same results, now, 
based on a different gauge choice which simplifies both the analyses 
and the results.
Results can be easily translated into solutions in any other gauge.

We consider a homogeneous, isotropic and flat cosmological model with 
general scalar type metric perturbations (we set $c \equiv 1$)
\bea
   d s^2 
   &=& - \left( 1 + 2 \alpha \right) d t^2
       - a \beta_{,\alpha} d t d x^\alpha
   \nonumber \\
   & & + a^2 \left[ \delta_{\alpha\beta} \left( 1 + 2 \varphi \right)
       + 2 \gamma_{,\alpha\beta} \right] d x^\alpha d x^\beta,
   \label{metric}
\eea
where $\alpha ({\bf x}, t)$, $\beta ({\bf x}, t)$, $\varphi ({\bf x}, t)$,
and $\gamma ({\bf x}, t)$ are perturbed order metric variables.
We consider the model is supported by a coherently oscillating minimally 
coupled scalar field with $V = {1\over 2} m^2 \phi^2$.
The scale we are interested in is
\bea
   & & m \gg k_p \gg H,
   \label{scales}
\eea
where $k_p \equiv k/a$ and $H \equiv \dot a/a$.
We have:
\bea
   & & {m \over H} = 4.689 \times 10^{27} h^{-1} \left( { m \over 
       10^{-5} {\rm eV}} \right) \left( {t \over t_0} \right),
   \nonumber \\
   & & \Gamma \equiv {k_p^2 \over m H} = 1.917 \times 10^{-15} 
   \nonumber \\
   & & \qquad \times
       \left( {m \over 10^{-5} {\rm eV}} \right)^{-1} 
       \left( {1 {\rm Kpc} \over k_{p 0}^{-1} } \right)^2
       \left( {t \over t_0} \right)^{-1/3},
\eea
where $H_0 \equiv 100 h {\rm km/ sec Mpc}$ with a subindex $0$ indicating the
present epoch.
In the following analysis we will strictly {\it ignore} 
$H/m$ order higher terms.
However, we will consider the {\it expansion} in $\Gamma \equiv k_p^2/(mH)$.

%%%%%%%%%%%%%%%%%%%%%%%%%%%%%%%%%%%%%%%%%%%%%%%%%%%%%%%%%%%%%%%%%%%%
{\it 2. Background evolution:}
Since the scalar field oscillates with the frequency $m$ which is
much larger than the characteristic expansion rate of the universe $H$,  
the temporal averages of the oscillating scalar field will contribute
to the background fluid quantities, and thus to the metric.
The set of equations describing the background evolution with a general 
minimally coupled scalar field is given in Eqs. (5-6) of \cite{H-MSF}.
Considering the time averaging we have:
\bea
   & & \mu = {1 \over 2} \langle \dot \phi^2 + m^2 \phi^2 \rangle, \quad
       p = {1 \over 2} \langle \dot \phi^2 - m^2 \phi^2 \rangle,
   \nonumber \\
   & & H^2 = {4 \pi G \over 3} \langle \dot \phi^2 + m^2 \phi^2 \rangle, \quad
       \dot H = - 4 \pi G \langle \dot \phi^2 \rangle, 
   \nonumber \\
   & & \ddot \phi + 3 H \dot \phi + m^2 \phi = 0.
   \label{BG-eqs} 
\eea
An angular bracket indicates an averaging over time scale of order $m^{-1}$.
Ignoring $H/m$ order higher terms we have an approximate solution,
\cite{Ratra}
\bea
   & & \phi (t) = a^{-3/2} \Big[ \phi_{+0} \sin{(mt)} 
       + \phi_{-0} \cos{(mt)} \Big],
   \label{BG-phi}
\eea
where $\phi_{+0}$ and $\phi_{-0}$ are constant coefficients.
Equation (\ref{BG-eqs}) becomes:
\bea
   & & \mu = \langle m^2 \phi^2 \rangle = \langle \dot \phi^2 \rangle
       = {1\over 2} m^2 a^{-3} \left( \phi_{+0}^2 + \phi_{-0}^2 \right),
   \nonumber \\
   & & p = 0, \quad
       \langle \phi \dot \phi \rangle = 0,
   \nonumber \\
   & & \dot H = - {3 \over 2} H^2, \quad H^2 = {8 \pi G \over 3} \mu, \quad
       a \propto t^{2/3}, \quad H = {2 \over 3 t}.
\eea
Thus, the background medium behaves exactly like a pressureless ideal fluid
\cite{Axion-CDM}.

%%%%%%%%%%%%%%%%%%%%%%%%%%%%%%%%%%%%%%%%%%%%%%%%%%%%%%%%%%%%%%%%%%%%
{\it 3. Perturbed equations in a gauge ready form:}
A complete set of equations describing the perturbed evolution of a
minimally coupled scalar field with a general potential is presented 
in a gauge ready form in \cite{H-MSF}.
In our case, we should take into account of the fact that due to the rapid
oscillation of the background scalar field, the perturbed part of the 
scalar field also oscillates, thus the properly time averaged quantities 
contribute to the evolution of the perturbed fluid quantities and metric.
Perturbed fluid quantities are presented in Eqs. (7) of \cite{H-MSF}:
\bea
   & & \varepsilon = \langle \dot \phi \delta \dot \phi - \dot \phi^2 \alpha 
       + m^2 \phi \delta \phi \rangle, \quad
   \nonumber \\
   & & \pi = \langle \dot \phi \delta \dot \phi - \dot \phi^2 \alpha 
       - m^2 \phi \delta \phi \rangle, 
   \nonumber \\
   & & \Psi = - \langle \dot \phi \delta \phi \rangle, \quad \sigma = 0. 
   \label{fluid-quantities}
\eea
Perturbed set of equations in a gauge ready form is presented in
Eqs. (8-13) of \cite{H-MSF}.
Considering the time averaging we have:
\bea
   & & \kappa = - 3 \dot \varphi + 3 H \alpha + {k^2 \over a^2} \chi, 
   \label{1} \\
   & & - H \kappa + {k^2 \over a^2} \varphi = 4 \pi G 
       \langle \dot \phi \delta \dot \phi - \dot \phi^2 \alpha 
       + m^2 \phi \delta \phi \rangle,
   \label{2} \\
   & & \kappa - {k^2 \over a^2} \chi = 12 \pi G
       \langle \dot \phi \delta \phi \rangle,
   \label{3} \\
   & & \alpha + \varphi = \dot \chi + H \chi, 
   \label{4} \\
   & & \dot \kappa + 2 H \kappa = \left( {k^2 \over a^2} 
       - 4 \pi G \langle \dot \phi^2 \rangle \right) \alpha 
   \nonumber \\
   & & \qquad \qquad \qquad \qquad
       + 8 \pi G
       \langle 2 \dot \phi \delta \dot \phi - m^2 \phi \delta \phi \rangle, 
   \label{5} \\
   & & \delta \ddot \phi + 3 H \delta \dot \phi + \left( {k^2 \over a^2}
       + m^2 \right) \delta \phi 
       = \dot \phi \left( \kappa + \dot \alpha \right)
   \nonumber \\
   & & \qquad \qquad \qquad \qquad
       - \left( 3 H \dot \phi + 2 m^2 \phi \right) \alpha.
   \label{6}
\eea
$\chi \equiv a (\beta + a \dot \gamma)$ is a spatially gauge invariant 
combination.
Every perturbed order variable in Eqs. (\ref{fluid-quantities}-\ref{6}) 
is spatially gauge-invariant, but depends on the temporal gauge condition.
As a gauge fixing condition we can impose one condition in any of these 
temporally gauge dependent variables.
Imposing any one of the following conditions is fine as a gauge condition:
$\alpha \equiv 0$ (synchronous gauge),
$\Psi \equiv 0$ (comoving gauge),
$\delta \phi \equiv 0$ (uniform-field gauge),
$\chi \equiv 0$ (zero-shear gauge),
$\varphi \equiv 0$ (uniform-curvature gauge),
$\kappa \equiv 0$ (uniform-expansion gauge),
$\varepsilon \equiv 0$ (uniform-density gauge), etc.
The decision is up to us.
Often the choice is made based on the author's taste. 
However, rather naturally, depending on the problem, the analysis in a
certain gauge condition is more convenient than in the other.
Sometimes, the advantage we get from the suitable gauge condition is exclusive.
By writing the equations without choosing the gauge as in 
Eqs. (\ref{1}-\ref{6}) we can try different gauge conditions as we wish.
We call this approach, which allows us to use the gauge issue not as a problem
but as an advantage, a gauge ready method. 
The solutions in other gauge conditions can be derived from the known 
solutions in a gauge.
Except for the synchronous gauge, any of other gauge conditions fixes
the gauge mode completely. 
Thus, every variable in such a gauge has the corresponding gauge invariant 
combination.
In this sense, the variables in such gauge conditions can be regarded as 
the equivalently gauge invariant ones.
Our experience tells that the uniform-curvature gauge is convenient for
problems involving the scalar field.
In the following we will start by adopting the uniform-curvature gauge,
or equivalently gauge invariant combinations.

%%%%%%%%%%%%%%%%%%%%%%%%%%%%%%%%%%%%%%%%%%%%%%%%%%%%%%%%%%%%%%%%%%%%
{\it 4. Analyses in the uniform-curvature gauge:}
We impose the uniform-curvature gauge condition which takes $\varphi \equiv 0$.
A gauge invariant combination is
\bea
   & & \delta \phi_\varphi \equiv \delta \phi - {\dot \phi \over H} \varphi,
   \label{UCG-UFG}
\eea
which becomes (thus, equivalent to) $\delta \phi$ in the uniform-curvature 
gauge.
{}For $\delta \phi_\varphi$ we take an {\it ansatz}, \cite{Ratra}
\bea
   & & \delta \phi_\varphi ({\bf k}, t) 
       = \delta \phi_{+} ({\bf k}, t) \sin{(mt)} 
       + \delta \phi_{-} ({\bf k}, t) \cos{(mt)}.
   \label{ansatz}
\eea
In the analyses of the perturbed order equations we
consider only the leading order terms in $H/m$.
However, since we are considering the small scale limit compared with the
horizon, we include higher order terms in the expansion of $\Gamma$.
{}From Eqs. (\ref{1},\ref{3}) we have
\bea
   & & \alpha_\varphi = {4 \pi G \over H} \langle \dot \phi 
       \delta \phi_\varphi \rangle.
   \label{UCG-alpha-eq}
\eea
Using Eqs. (\ref{2},\ref{UCG-alpha-eq}), Eq. (\ref{6}) leads to
\bea
   & & \delta \ddot \phi_\varphi + 3 H \delta \dot \phi_\varphi
       + \left( {k^2 \over a^2} + m^2 \right) \delta \phi_\varphi
   \nonumber \\
   & & \qquad
       = - {8 \pi G \over H} m^2 \Big( 
       \phi \langle \dot \phi \delta \phi_\varphi \rangle
       + \dot \phi \langle \phi \delta \phi_\varphi \rangle \Big)
   \nonumber \\
   & & \qquad
       = - {3 H \over \phi_{+0}^2 + \phi_{-0}^2 } a^{3/2}
       \Big[ m \phi \left( \phi_{+0} \delta \phi_{-} 
       - \phi_{-0} \delta \phi_{+} \right)
   \nonumber \\
   & & \qquad \qquad
       + \dot \phi \left( \phi_{+0} \delta \phi_{+}
       + \phi_{-0} \delta \phi_{-} \right) \Big],
   \label{UCG-delta-phi-1}
\eea
where in the second step we used Eqs. (\ref{BG-phi},\ref{ansatz}).
Using Eqs. (\ref{BG-phi},\ref{ansatz}), Eq. (\ref{UCG-delta-phi-1}) leads to:
\bea
   & & t \delta \dot \phi_{-} + { 2 \phi_{-0}^2 \over \phi_{+0}^2 
       + \phi_{-0}^2 } \delta \phi_{-} 
       = \left( - { 2 \phi_{+0} \phi_{-0} \over \phi_{+0}^2 + \phi_{-0}^2 } 
       + {1\over 3} \Gamma \right) \delta \phi_{+},
   \nonumber \\
   & & t \delta \dot \phi_{+} + { 2 \phi_{+0}^2 \over \phi_{+0}^2 
       + \phi_{-0}^2 } \delta \phi_{+} 
       = \left( - { 2 \phi_{+0} \phi_{-0} \over \phi_{+0}^2 + \phi_{-0}^2 } 
       - {1\over 3} \Gamma \right) \delta \phi_{-}.
   \nonumber \\
\eea
The solution valid to second order in the {\it expansion} of $\Gamma$ is
(notice that $\Gamma \propto t^{-1/3}$)
\bea
   \delta \phi_{+} ({\bf k}, t) 
   &=& c_1 ({\bf k}) \left( 1 + {1\over 5} {\phi_{+0} \over \phi_{-0} }
       \Gamma + {1\over 10} \Gamma^2 \right)
   \nonumber \\
   & & + c_2 ({\bf k}) t^{-2} \left( 1 + {1\over 7} {\phi_{-0} \over \phi_{+0} }
       \Gamma - {1\over 14} \Gamma^2 \right),
   \nonumber \\
   \delta \phi_{-} ({\bf k}, t) 
   &=& c_1 ({\bf k}) \left( - {\phi_{+0} \over \phi_{-0} } 
       + {1\over 5} \Gamma
       - {1\over 10} {\phi_{+0} \over \phi_{-0} } \Gamma^2 \right)
   \nonumber \\
   & & + c_2 ({\bf k}) t^{-2} \left( {\phi_{-0} \over \phi_{+0} } 
       - {1\over 7} \Gamma
       - {1\over 14} {\phi_{-0} \over \phi_{+0} } \Gamma^2 \right).
   \label{UCG-phi-sol1}
\eea
By the following identification of the coefficients
\bea
   C ({\bf k})
   &\equiv& {2 \over 3} \left( { a^{3/2} t^{-1} \over \phi_{-0} } \right)
       {1\over m} c_1 ({\bf k}),
   \nonumber \\
   d ({\bf k})
   &\equiv& {27 \over 4} \left( { a^{3/2} t^{-1} \over \phi_{+0} } \right)
       \left( {H \over k_p} t^{1/3} \right)^2 c_2 ({\bf k}),
       \label{UCG-matching}
\eea
Eqs.(\ref{ansatz},\ref{UCG-phi-sol1}) lead to
\bea
   & & \delta \phi_\varphi ({\bf k}, t) 
       = - C ({\bf k}) {1 \over H}
       \Big( \dot \phi - {1\over 5} m \phi \Gamma
       + {1\over 10} \dot \phi \Gamma^2 \Big)
   \nonumber \\
   & & \qquad
       + {4 \over 27} d ({\bf k}) t^{-5/3} \left( {k_p \over H} \right)^2
       \Big( \phi - {1\over 7} {\dot \phi \over m} \Gamma
       - {1\over 14} \phi \Gamma^2 \Big).
   \label{UCG-delta-phi}
\eea
In order to derive the complete set of solutions in the other gauge conditions 
we need solutions valid to second order in the expansion of $\Gamma$.
Using the solution in Eq. (\ref{UCG-delta-phi}) the other perturbed 
quantities follow from Eqs. (\ref{1}-\ref{5}) as:
\bea
   & & \alpha_\varphi = - {3 \over 2} {H \over \mu} \Psi_\varphi
       = - {3 \over 2} C \Big( 1 + {1\over 10} \Gamma^2 \Big)
   \nonumber \\
   & & \qquad
       - {2 \over 63} d t^{-5/3} \Gamma^2 \Big( 1 + 0 \cdot \Gamma \Big),
   \nonumber \\
   & & H \chi_\varphi = \left( {H \over k_p} \right)^2 {\kappa_\varphi \over H}
       = - {3 \over 2} \left( {H \over k_p} \right)^2 \delta_\varphi
   \nonumber \\
   & & \qquad
       = - {3 \over 5} C 
       \Big( 1 + 0 \cdot \Gamma \Big)
       - {4 \over 9} d t^{-5/3} \Big( 1 - {1\over 14} \Gamma^2 \Big).
   \label{UCG-sols} 
\eea

In the pressureless limit the relativistic cosmological perturbation
of an ideal fluid reduces to the Newtonian one.
Newtonian hydrodynamics is described by the following quantities:
the density, pressure, velocity, and gravitational potential.
In Sec. 4 of \cite{H-MDE} we made some arguments that in the Einstein gravity 
filled with an ideal fluid, the following gauge invariant combinations 
play the roles of the Newtonian relative density fluctuation 
($\delta \varrho / \varrho$), potential fluctuation ($- \delta \Phi$),
and velocity fluctuation ($\delta v$), respectively:
\bea
   & & \delta_\Psi \equiv {\varepsilon - 3 H \Psi \over \mu}, \quad
       \varphi_\chi \equiv \varphi - H \chi,
   \nonumber \\
   & & v_\chi \equiv - {k \over a} {\Psi + (\mu + p) \chi \over \mu + p}.
\eea
$\delta_\Psi$, $\varphi_\chi$, and $v_\chi$ are the relative density 
fluctuation in the comoving gauge, the potential fluctuation in the
zero-shear gauge, and the velocity fluctuation in the zero-shear gauge, 
respectively \cite{Bardeen-1980}.
In our case, from Eq. (\ref{UCG-sols}), the gauge invariant combinations become:
\bea
   & & \delta_\Psi 
       = \left( {k_p \over H} \right)^2 \Bigg[ {2\over 5} C 
       \Big( 1 + 0 \cdot \Gamma \Big)
       + {8 \over 27} d t^{-5/3} \Big( 
       1 - {1\over 14} \Gamma^2 \Big) \Bigg],
   \nonumber \\
   & & \varphi_\chi
       = {3\over 5} C \Big( 1 + 0 \cdot \Gamma \Big)
       + {4 \over 9} d t^{-5/3} \Big( 1 - {1\over 14} \Gamma^2 \Big),
   \nonumber \\
   & & v_\chi 
       = {k_p \over H} \Bigg[ - {2\over 5} C \Big( 1 + 0 \cdot \Gamma \Big)
       + {4 \over 9} d t^{-5/3} \Big( 1 - {5\over 42} \Gamma^2 \Big) 
       \Bigg].
   \nonumber \\
   \label{v-chi}
\eea
On the other hand, a thorough study of the evolution of perturbations in
a pressureless ideal fluid was made in \cite{H-MDE}.
{}From Table 2 of \cite{H-MDE} we have (these solutions are valid in general
scales):
\bea
   & & \delta_\Psi 
       = \left( {k_p \over H} t^{-1/3} \right)^2
       \left( {2\over 5} C t^{2/3} + {8 \over 27} d t^{-1} \right),
   \nonumber \\
   & & \varphi_\chi 
       = {3\over 5} C + {4 \over 9} d t^{-5/3},
   \nonumber \\
   & & v_\chi
       = \left( {k_p \over H} t^{-1/3} \right)
       \left( - {2 \over 5} C t^{1/3} + {4 \over 9} d t^{-4/3} \right).
   \label{v-chi-IF}
\eea
Thus, to the linear order in $\Gamma$ Eq. (\ref{v-chi})
coincides with Eq. (\ref{v-chi-IF}).
Therefore, the perturbed part of the coherently oscillating scalar field 
behaves as a perturbed pressureless fluid, thus as a cold dark matter.

%%%%%%%%%%%%%%%%%%%%%%%%%%%%%%%%%%%%%%%%%%%%%%%%%%%%%%%%%%%%%%%%%%%%
{\it 5. Comparison with the general solutions in the large-scale limit:}
It is interesting to compare the solutions in 
Eqs. (\ref{UCG-delta-phi},\ref{UCG-sols}) with the general integral form
solutions valid in the large scale limit ($k_p \ll H$).
A general equation for $\delta \phi_\varphi$ is derived in \cite{H-QFT} as
\bea
   & & \delta \ddot \phi_\varphi + 3 H \delta \dot \phi_\varphi
       + \Bigg[ {k^2 \over a^2} 
   \nonumber \\
   & & \qquad
       + V_{,\phi\phi} + 2 {\dot H \over H}
       \left( 3 H - {\dot H \over H} + 2 {\ddot \phi \over \dot \phi} \right)
       \Bigg] \delta \phi_\varphi = 0.
   \label{MSF-eq}
\eea
This equation can be derived from Eqs. (\ref{1}-\ref{3},\ref{6},\ref{BG-eqs}) 
for a general minimally coupled scalar field without the coherent oscillation.
If we consider the coherent oscillation and the corresponding contribution 
of the time averaged quantities to the metric, the same equations instead 
will lead to Eq. (\ref{UCG-delta-phi-1}).
{}For $k_p \ll H$ we have
\bea
   & & \delta \phi_\varphi ({\bf x},t)
       = - {\dot \phi \over H} \left[ C ({\bf x})
       - D ({\bf x}) \int^t_0 {H^2 \over a^3 \dot \phi^2 } dt \right].
   \label{MSF-LS-sol}
\eea
Equations (\ref{MSF-eq},\ref{MSF-LS-sol}) are valid for an arbitrary $V(\phi)$. 
In Eq. (109) of \cite{GGT-HN} we proved that \cite{COMMENT-1} 
\bea
   & & D ({\bf k}) = {4 \over 3} k^2 \left( {a \over t^{2/3}} \right) 
       d ({\bf k}).
   \label{D-d}
\eea
Thus, in the large scale limit the growing mode of Eq. (\ref{UCG-delta-phi}) 
coincides with the one in Eq. (\ref{MSF-LS-sol}).
A complete set of the large scale asymptotic solutions is presented in
\cite{H-MSF}.
{}From the Table 1 \cite{H-MSF} we have \cite{COMMENT-2}:
\bea
   & & {H \over \mu} \Psi_\varphi = - {2 \over 3} \alpha_\varphi
       = - {H \over \dot \phi} \delta \phi_\varphi
       = {1 \over 3} \delta_\varphi
       = - {2 \over 9} {\kappa_\varphi \over H} = C,
   \nonumber \\
   & & H \chi_\varphi = - {3 \over 5} C - \bar d {H \over a}.
   \label{UCG-sols-IF}
\eea
The second term in $\chi_\varphi$ corresponds to a decaying mode.
The growing modes of $\Psi_\varphi$, $\alpha_\varphi$, $\delta \phi_\varphi$, 
and $\chi_\varphi$
in Eq. (\ref{UCG-sols-IF}) coincide with the ones in Eq. (\ref{UCG-sols}).
However, behaviors of $\delta_\varphi$ and $\kappa_\varphi$ are different.
Since we have derived Eqs. (\ref{UCG-delta-phi},\ref{UCG-sols}) based on 
Eq. (\ref{scales}) they are not necessarily valid in the $k_p \ll H$ limit.
Never the less, from Eq. (\ref{UCG-sols-IF}) we can show that the gauge 
invariant combinations in Eq. (\ref{v-chi}) {\it remain valid} 
in the large scale limit; a proof in a more general ground can be found 
by comparing the large scale solutions presented in the Table 1 of \cite{H-MSF}
with the exact solutions valid in the general scale
presented in the Table 1 of \cite{H-MDE}.
Therefore, Eqs. (\ref{v-chi},\ref{v-chi-IF}) are also valid in the
superhorizon scale, which implies that, in this sense,
the coherently oscillating scalar field behaves as a
cold dark matter on general scales
as long as we have $\Gamma \ll 1$ (and $H/m \ll 1$).

%%%%%%%%%%%%%%%%%%%%%%%%%%%%%%%%%%%%%%%%%%%%%%%%%%%%%%%%%%%%%%%%%%%%
{\it 6. Solutions in the other gauges:}
{}From the complete set of solutions derived in the uniform-curvature gauge
[Eqs. (\ref{UCG-delta-phi},\ref{UCG-sols})] we can derive the rest of the 
solutions in the other gauge conditions.
This translation into other gauges can be done systematically using either
the gauge transformation or various gauge invariant combinations; the
latter method is much easier as shown in deriving Eq. (\ref{v-chi}). 
In practice, it is convenient to derive any one variable in the gauge we are 
interested using any of the above methods, and then, to derive the rest of the 
perturbed metric and fluid variables using the fundamental set of equations
in Eqs. (\ref{1}-\ref{6}).
In the {\it zero-shear gauge} we set $\chi \equiv 0$.
{}From Eqs. (17,19) of \cite{H-PRW} we have
$\delta \phi_\chi \equiv \delta \phi - \dot \phi \chi
 = \delta \phi_\varphi - \dot \phi \chi_\varphi$.
Thus, from Eqs. (\ref{UCG-delta-phi},\ref{UCG-sols}) we have
\bea
   \delta \phi_\chi 
   &=& - {2 \over 5 H} C \Big( \dot \phi
       - {1 \over 2} m \phi \Gamma \Big)
   \nonumber \\
   & & + {2\over 3} d t^{-2/3} \Big( \dot \phi + {1\over 3} m \phi \Gamma
       - {5 \over 42} \dot \phi \Gamma^2 \Big).
   \label{ZSG-delta-phi}
\eea
The rest of the perturbed variables follow from Eqs. (\ref{1}-\ref{6}).
The {\it uniform-expansion gauge} takes $\kappa = 0$.
{}From Eqs. (17,19) of \cite{H-PRW} we have
$\delta \phi_\kappa \equiv \delta \phi + \dot \phi
 (3 \dot H - k_p^2)^{-1} \kappa$.
Since $H^2/k_p^2 = (H/m)/\Gamma$, we neglect
$H^2$ term compared with $k_p^2$ one. 
{}From Eqs. (\ref{UCG-sols}) we have $\chi_\varphi = \kappa_\varphi/ k_p^2$, 
thus we have $\delta \phi_\kappa = \delta \phi - \dot \phi (\kappa/k_p^2)
 = \delta \phi - \dot \phi \chi = \delta \phi_\chi$.
Thus, the solution in Eq. (\ref{ZSG-delta-phi}) is also valid in 
the uniform-expansion gauge.
The {\it synchronous gauge} takes $\alpha = 0$.
{}From Eqs. (17,19) of \cite{H-PRW} we have
$\delta \phi_\alpha \equiv \delta \phi - \dot \phi \int^t \alpha dt$.
The combination with a subindex $\alpha$ is not gauge invariant.
The lower bound of the integration in the right hand side leads to the
behavior of the gauge mode.
{}From Eqs. (\ref{UCG-delta-phi},\ref{UCG-sols}) we can derive
\bea
   & & \delta \phi_\alpha 
       = \left( {k_p \over H} \right)^2
       \Bigg[ {1 \over 5} C \Big( \phi + {\dot \phi \over m} \Gamma \Big)
   \nonumber \\
   & & \quad
       + {4\over 27} d t^{-5/3} \Big( \phi
       - {1\over 4} {\dot \phi \over m} \Gamma
       - {1\over 14} \phi \Gamma^2 \Big) \Bigg]
       + {\rm const.} \times \dot \phi,
\eea
where the last term is the gauge mode.

%%%%%%%%%%%%%%%%%%%%%%%%%%%%%%%%%%%%%%%%%%%%%%%%%%%%%%%%%%%%%%
{\it 7. Discussions:}
When we have $H/m \ll 1$ the homogeneous part of a coherently oscillating 
scalar field behaves as a pressureless medium (\S 2).
In such a background, we have shown that on scales 
satisfying $\Gamma \ll 1$ the perturbed part of a coherently oscillating 
scalar field behaves as a cold dark matter (\S 4).
We have shown that this conclusion is valid even for scales larger 
than the horizon (\S 5).
We followed the method suggested in \cite{Ratra}, but took the 
uniform-curvature gauge which allows simpler analysis. 
Solutions in any other gauge can be derived easily (\S 6).
We expect the equations and results presented above will be useful
for handling the reheating process based on the coherently oscillating 
scalar field \cite{reheating}.
Applications to the reheating process will be presented elsewhere.

%%%%%%%%%%%%%%%%%%%%%%%%%%%%%%%%%%%%%%%%%%%%%%%%%%%%%%%%%%%%%%
We thank Dr. S.-J. Rey for useful discussions and encouragement.
We also wish to thank Drs. H. Noh and D. Ryu for useful suggestions.

%%%%%%%%%%%%%%%%%%%%%%%%%%%%%%%%%%%%%%%%%%%%%%%%%%%%%%%%%%%%%%%%%

%%%%%%%%%%%%%%%%%%%%%%%%%%%%%%%%%%%%%%%%%%%%%%%%%%%%%%%%%%%%%%%%%

%%%%%%%%%%%%%%%%%%%%%%%%%%%%%%%%%%%%%%%%%%%%%%%%%%%%%%%%%%%%%%%%%

\begin{references}
\bibitem{Axion-CDM}
         L. F. Abbott and P. Sikivie, Phys. Lett. B {\bf 120}, 133 (1983);
         M. Axenides, R. Brandenberger and M. Turner, Phys. Lett. {\bf 126}, 
            178 (1983);
         M. Dine and W. Fischler, Phys. Lett. B {\bf 120}, 137 (1983);
         M. Fukugita and M. Yoshimura, Phys. Lett. B {\bf 127}, 181 (1983);
         J. Ipser and P. Sikivie, Phys. Rev. Lett. {\bf 50}, 925 (1983);
         A. L. Melott, {\it et. al.}, Phys. Rev. Lett. {\bf 51}, 935 (1983);
         J. Preskill, M. B. Wise and F. Wilczek, Phys. Lett. B {\bf 120}, 
            127 (1983);
         F. W. Stecker and Q. Shafi, Phys. Rev. Lett. {\bf 50}, 1128 (1983);
         P. J. Steinhardt and M. S. Turner, Phys. Lett. B {\bf 74}, 3105 (1983);
         M. S. Turner, Phys. Rev. D {\bf 28}, 1243 (1983);
         M. S. Turner, F. Wilczek and A. Zee, Phys. Lett. B {\bf 125}, 35 
               (1983);
         L. F. Abbott and M. B. Wise, Nucl. Phys. B {\bf 237}, 226 (1984);
         Q. Shafi and F. W. Stecker, Phys. Rev. Lett. {\bf 53}, 1292 (1984);
         D. Seckel and M. S. Turner, Phys. Rev. D {\bf 32}, 3178 (1985);
         J. E. Kim, Phys. Rep. {\bf 150}, 1 (1987);
         C. J. Hogan and M. J. Rees, Phys. Lett. B {\bf 205}, 228 (1988);
         Y. Nambu and M. Sasaki, Phys. Rev. D {\bf 42}, 3918 (1990);
         M. S. Turner, Phys. Rep. {\bf 197}, 67 (1990);
         M. S. Turner and F. Wilczek, Phys. Rev. Lett. {\bf 66}, 5 (1991).
\bibitem{Brandenberger}
         R. H. Brandenberger, Phys. Rev. D {\bf 32}, 501 (1984).
\bibitem{Sasaki}
         M. Sasaki, Prog. Theor. Phys. {\bf 72}, 1266 (1984);
         H. Kodama and M. Sasaki, Prog. Theor. Phys. Suppl. {\bf 78}, 1 (1984).
\bibitem{Ratra}
         B. Ratra, Phys. Rev. D {\bf 44}, 352 (1991).
\bibitem{H-MSF}
         J. Hwang, Astrophys. J. {\bf 427}, 542 (1994).
\bibitem{H-MDE}
         J. Hwang, Astrophys. J. {\bf 427}, 533 (1994).
\bibitem{Bardeen-1980}
         J. M. Bardeen, Phys. Rev. D {\bf 22}, 1882 (1980).
\bibitem{H-QFT}
         J. Hwang, Phys. Rev. D {\bf 48}, 3544 (1993).
\bibitem{GGT-HN}
         J. Hwang and H. Noh, Phys. Rev. D {\bf 54}, 1460 (1996).
\bibitem{COMMENT-1}
         We note that $d$ used in \cite{GGT-HN}, let's denote it as $\bar d$,
         differs from our $d$ as $\bar d = {2 \over 3} (a/t^{2/3}) d$.
\bibitem{COMMENT-2}
         As shown in Eq. (\ref{D-d}) the decaying mode in 
         Eq. (\ref{MSF-LS-sol}) is higher order in the large scale expansion 
         compared with the solutions in the other gauges, 
         thus disappearing in Eq. (\ref{UCG-sols-IF}) which is
         valid in the large scale limit. 
\bibitem{H-PRW}
         J. Hwang, Astrophys. J. {\bf 375}, 443 (1991).
\bibitem{reheating}
         J. Traschen and R. Brandenberger, Phys. Rev. D {\bf 42}, 2491 (1990);
         L. Kofman, A. Linde and A. A. Starobinsky, Phys. Rev. Lett.
            {\bf 73}, 1425 (1994);
         Y. Shtanov, J. Traschen and R. Brandenberger, Phys. Rev. D
            {\bf 51}, 5438 (1995);
         M. Yoshimura, Prog. Theor. Phys. {\bf 94}, 873 (1995);
         Y. Nambu and A. Taruya, gr-qc/9606029;
         H. Kodama and T. Hamazaki, gr-qc/9608022;
         T. Hamazaki and H. Kodama, gr-qc/9609036.
\end{references}
\end{document}